\documentclass[reprint, superscriptaddress, secnumarabic, amssymb, nobibnotes, aps, prl]{revtex4-1}

\usepackage{graphicx}
\usepackage{epstopdf}
\usepackage[T1]{fontenc}
\usepackage[latin9]{inputenc}
\usepackage{amsbsy}
\usepackage{gensymb}
\usepackage{textcomp}
\usepackage[T1]{fontenc}
\usepackage[latin9]{inputenc}
\usepackage{amsmath}
\usepackage{amssymb}
\usepackage{bbm}
\usepackage{soul}
\usepackage{physics}
\usepackage{braket}
\usepackage{xcolor}
\allowdisplaybreaks
\usepackage{graphicx}
\usepackage[colorlinks=true]{hyperref}  
\hypersetup{
    unicode=false,          
    pdftoolbar=true,        
    pdfmenubar=true,        
    pdffitwindow=false,     
    pdfstartview={FitH},    
    pdftitle={Superconductivity in Equimolar High Entropy Alloy \texorpdfstring{Nb$_{0.20}$Mo$_{0.20}$Re$_{0.20}$Ru$_{0.20}$Ir$_{0.20}$}{}},    
    pdfauthor={, , ,},     
    pdfsubject={},   
    pdfcreator={},   
    pdfproducer={},  
    pdfkeywords={} {} {}, 
    pdfnewwindow=true,      
    colorlinks=true,       
    linkcolor=blue, 
    citecolor=blue,        
    filecolor=magenta,      
    urlcolor=blue           
}
\usepackage[normalem]{ulem}

\newcommand{\figref}[1]{Fig.~\ref{#1}}

\begin{document}
\title{High critical field superconductivity in a 3d dominated lightweight equiatomic high entropy alloy}

\author{S. Jangid}
\affiliation{Department of Physics, Indian Institute of Science Education and Research Bhopal, Bhopal, 462066, India}
\author{P. K. Meena}
\affiliation{Department of Physics, Indian Institute of Science Education and Research Bhopal, Bhopal, 462066, India}
\author{R. K. Kushwaha}
\affiliation{Department of Physics, Indian Institute of Science Education and Research Bhopal, Bhopal, 462066, India}
\author{S. Srivastava}
\affiliation{Department of Physics, Indian Institute of Science Education and Research Bhopal, Bhopal, 462066, India}
\author{P. Manna}
\affiliation{Department of Physics, Indian Institute of Science Education and Research Bhopal, Bhopal, 462066, India}
\author{S. Sharma}
\affiliation{Department of Physics, Indian Institute of Science Education and Research Bhopal, Bhopal, 462066, India}
\author{P. Mishra}
\affiliation{Department of Physics, Indian Institute of Science Education and Research Bhopal, Bhopal, 462066, India}
\author{R.~P.~Singh}
\email[]{rpsingh@iiserb.ac.in}
\affiliation{Department of Physics, Indian Institute of Science Education and Research Bhopal, Bhopal, 462066, India}

\begin{abstract}
The lightweight high entropy alloy represents an innovative class of multicomponent systems that combine low density with the exceptional mechanical properties of high-entropy alloys. We present a detailed synthesis and investigation of a 3d rich equiatomic high entropy alloy superconductor Sc-Ti-V-Nb-Cu, which crystallizes in a body-centered cubic structure. Magnetization, electrical resistivity, and heat capacity measurements confirm weakly coupled bulk type II superconductivity with a 7.21(3) K transition temperature and an upper critical field of 12.9(1) T. The upper critical field approaches the Pauli paramagnetic limit, suggesting potential unconventional behavior. The low density, moderate transition temperature, and high upper critical field stand out Sc-Ti-V-Nb-Cu as a promising candidate for next-generation superconducting device applications.
\end{abstract}
\maketitle

High entropy alloys (HEAs) have emerged as a significant breakthrough in materials science, offering unique chemical complexity and outstanding mechanical properties \cite{yeh2004nanostructured}. These alloys are composed of five or more elements in significant ratios \cite{jien2006recent}, where each atom is encompassed by a number of different atoms, resulting in lattice strain and stress that induce significant lattice deformation \cite{Yeh2013}. This is generally recognized as the primary factor responsible for the extraordinary mechanical properties of HEA, such as high strength, superior thermal stability, remarkable resistance to oxidation and corrosion, and high fracture toughness at cryogenic temperatures \cite{gludovatz2014fracture, zou2015ultrastrong, lee2007effect}. Such attributes make them highly valuable for a wide range of applications, including structural and functional materials, energy storage, magnetic refrigeration, radiation protection, biocompatibility, and superconducting magnets in harsh environments \cite{funct, hstorage, radiation, biomat}.

Recently, superconducting high entropy alloys have emerged as a promising field of research, offering the opportunity to integrate the unique features of these alloys with the exceptional electrical and magnetic properties of superconductors \cite{kovzelj2014discovery}. These disordered alloys have shown exotic superconducting properties, such as the ability to retain superconductivity under high pressures, high critical current densities, elevated upper critical fields, broadening in heat capacity jump and elemental range of Debye temperature \cite{robustSCpressure, motla2023superconducting, jangid2024superconductivity, jc, heathinfilm, kasem2021anomalous, marik2019superconductivity, motla2021probing}. Extensive research on HEA superconductors has explored various aspects, including the correlation of $T_C$ and valence electron concentration (VEC), mechanical properties, electronic structure, microscopic investigations, isoelectronic substitutions, pressure effects, and fabrication techniques for thin films \cite{von2016effect, Kitagawa2022, jasiewicz2016, motla2021probing, motla2022superconducting, vonrohr2018, robustSCpressure, heathinfilm}. However, most HEAs and superconducting HEAs are based on refractory or rare earth elements, which significantly limit their potential applications \cite{kitagawa2020cutting}.

Lightweight high entropy alloys (LWHEAs), a subset of HEAs with densities below 7 g/cm$^3$ \cite{lightweightHEA}, have garnered significant attention for their potential application in the automotive and aerospace industries, where reducing weight without compromising structural integrity is crucial \cite{kumar2016insight}. A practical approach for developing LWHEAs involves the incorporation of low-density elements, especially 3d transition metals. Despite growing interest in lightweight materials, the development of superconducting LWHEAs remains unexplored. Such materials could revolutionize low-temperature superconducting devices, offering significant advantages in extreme environments and space applications due to their reduced weight and enhanced performance \cite{kumar2016insight}.

Herein, we report the synthesis and comprehensive study of the superconducting characteristics of a new 3d rich equiatomic LWHEA superconductor Sc$_{0.20}$Ti$_{0.20}$V$_{0.20}$Nb$_{0.20}$Cu$_{0.20}$. It crystallizes in a body-centered cubic (bcc) structure and exhibits weakly coupled type-II bulk superconductivity with a transition temperature of 7.21(3) K and an upper critical field of 12.9(1) T. The unique combination of equiatomic composition, majority of 3d elements, and high superconducting transition temperature and upper critical field makes this LWHEA a promising candidate for understanding superconducting pairing mechanisms and developing superconducting devices.
\begin{figure*}
\includegraphics[width=2.0\columnwidth, origin=b]{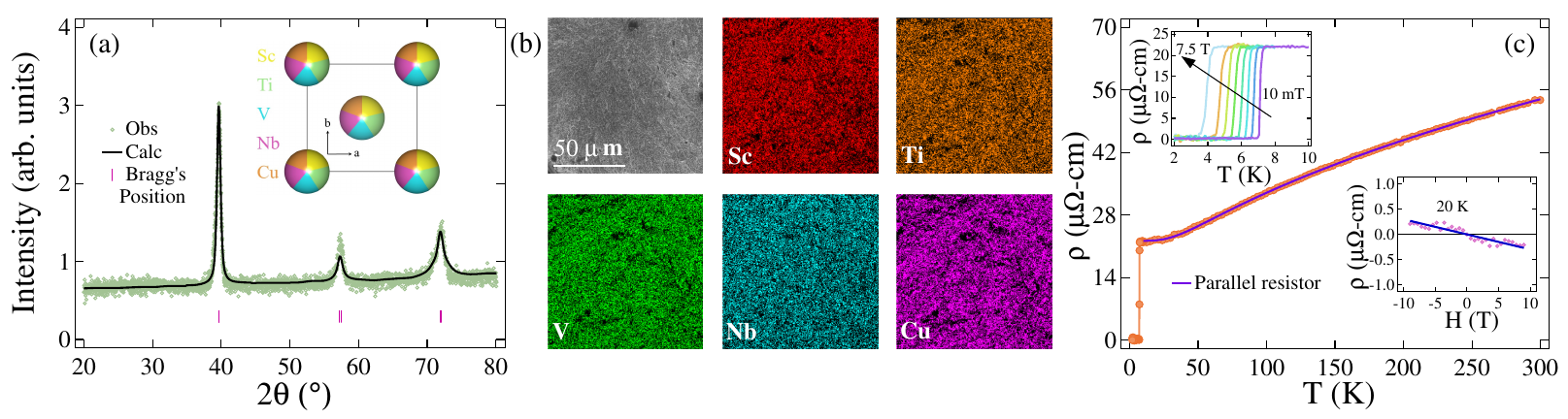}
\caption{\label{Fig1}(a) Powder X-ray diffraction pattern of Sc-Ti-V-Nb-Cu recorded at room temperature, which shows the crystallization in a bcc structure. The inset illustrates the atomic arrangement in the crystal structure of Sc-Ti-V-Nb-Cu. (b) SEM mapping of elements Sc, Ti, V, Nb, and Cu. (c) Electrical resistivity under a zero applied field in the temperature range of 1.9-300 K. The solid violet line represents the fit to the parallel resistor model. The top inset shows temperature variation of resistivity near the transition temperature under various applied fields, and the bottom inset shows the field variation of Hall resistivity from -9 T to 9 T at 20 K.}
\end{figure*}

A polycrystalline sample of Sc-Ti-V-Nb-Cu was synthesized using a conventional arc melter. High-purity (4N) elements Sc, Ti, V, Nb, and Cu were combined in the stoichiometric ratio and arc-melted under an argon atmosphere. In order to remove any residual oxygen, a Ti getter was melted prior to melting of the elements. The resultant ingot was melted several times by flipping it over to achieve phase homogeneity. There was negligible weight loss. The crystal structure and phase purity were confirmed by powder X-ray diffraction using a PANalytic X$^{'}$Pert diffractometer with CuK$_{\alpha}$ ($\lambda = 1.5406 \text{\AA}$) radiation. Energy-dispersive X-ray diffraction (EDX)was performed using a scanning tunneling microscope (SEM) to examine the phase composition and homogeneity. Magnetization measurements were carried out using a Quantum Design Magnetic Property Measurement System (MPMS 3) with a vibrating sample magnetometer. The electrical resistivity and specific heat measurements were performed using a Quantum Design Physical Property Measurement System (PPMS) via a four-probe and two-tau time-relaxation technique.


The powder X-ray diffraction pattern of Sc-Ti-V-Nb-Cu, as shown in \figref{Fig1}(a), confirms the adaptation of a body-centered cubic structure with the space group Im3$\bar{m}$ (229). HighScore Plus software was used to perform the Le-Bail refinement \cite{le1988ab}, which yields the lattice parameters $a=b=c=3.2141(4)$ \text{\AA}, cell volume $V_{cell}=33.20(4)$ \text{\AA}$^3$ and density $d=6.01(1)$ g/cm$^{3}$. The inset of \figref{Fig1}(a) shows the crystal structure of Sc-Ti-V-Nb-Cu with high site-mixing. The average phase composition was obtained to be Sc$_{0.23}$Ti$_{0.20}$V$_{0.19}$Nb$_{0.19}$Cu$_{0.19}$ by the EDX analysis performed at different sample locations, which is close to the nominal composition considering the experimental error. Furthermore, the uniform distribution of Sc, V, Ti, Cu, and Nb elements supports the phase homogeneity, as shown by the EDX elemental mapping in \figref{Fig1}(b).

\begin{figure*}[ht]
\includegraphics[width=2\columnwidth, origin=b]{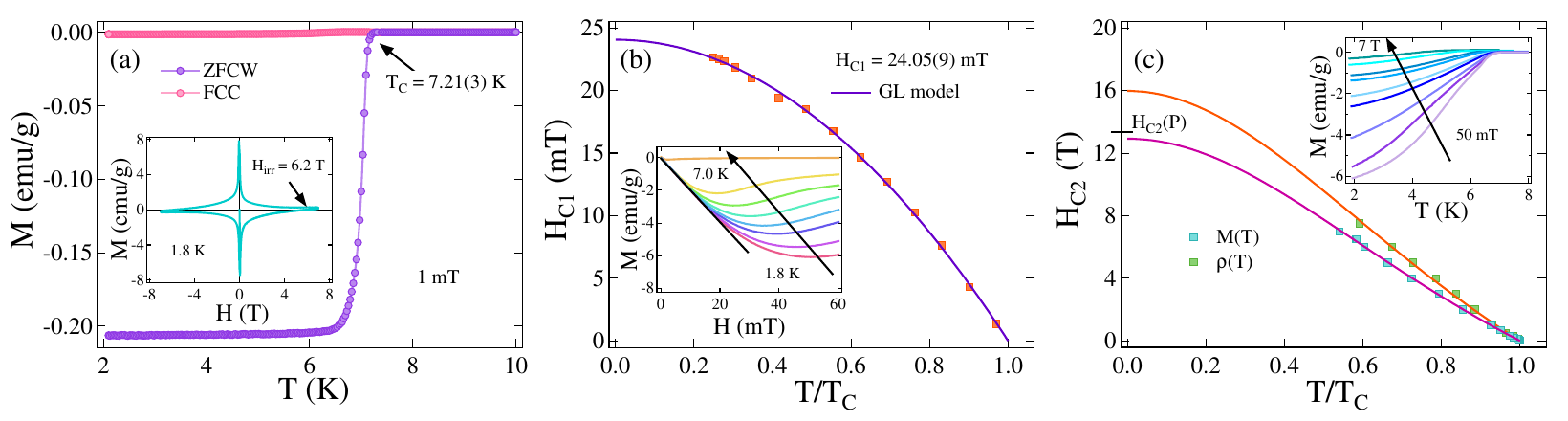}
\caption{\label{Fig2}(a) Temperature dependence of magnetization in ZFCW and FCC modes at 1.0 mT. The magnetization loop at 1.8 K, as shown in the inset, confirms type-II superconductivity in Sc-Ti-V-Nb-Cu. (b) Temperature variation of the lower critical field $H_{C1}(T)$, fitted with GL model, with inset showing isothermal magnetization curves recorded at various temperatures. (c) Temperature variation of the upper critical field $H_{C2}(T)$ calculated from magnetization (sky blue squares) and resistivity (green squares) measurements fitted with the GL model, and the inset shows temperature-dependent magnetization obtained at various applied fields.}
\end{figure*}
The temperature evolution of the zero field electrical resistivity $\rho(T)$ for Sc-Ti-V-Nb-Cu was measured from 1.9 to 300 K, as shown in \figref{Fig1}(c). At the transition temperature $T_C = 7.32(2)$, a zero drop in $\rho(T)$ is observed, confirming the superconductivity. The sluggish increase in the resistivity in the normal state with respect to temperature and the residual resistivity ratio (RRR) $\rho_{300 K}/\rho_{10 K}= 2.44(7)$ indicate the poor metallic nature of Sc-Ti-V-Nb-Cu. The electrical resistivity above $T_C$ is accurately described by the following expression of the parallel resistor model \cite{wiesmann1977simple}:
\begin{equation}
    \frac{1}{\rho(T)}=\frac{1}{\rho_1(T)}+\frac{1}{\rho_{sat}},
\end{equation}
where $\rho_{sat}$ represent the saturation resistivity whereas $\rho_1(T)$ is temperature dependent resistivity given by the following expression:
\begin{equation}
    \rho_1(T)=\rho_0+R\left(\frac{T}{\theta_{R}}\right)^5 \int^{\theta_{R}/T}_0 \frac{x^5}{\left(e^x-1\right)\left(1-e^{-x}\right)}\,dx.
    \label{eqn1:BG}
\end{equation}
Here, the first term $\rho_0$ represents the residual resistivity, and the subsequent term is the Bloch-Gr\"{u}neisen term \cite{Grimvall1981electron}, where $R$ is a material-dependent parameter and $\theta_{R}$ refers to the Debye temperature \cite{bid2006temperature}. The fitting of the data yields $\rho_0 = 26.19(9)\ \mu\Omega$-cm, $\rho_{sat} = 146(1)\ \mu\Omega$-cm and $\theta_{R}=197(2)$ K.
The Kadowaki-Woods ratio ($K_w = A/\gamma^2_n$) can be used to analyze electron-electron correlation \cite{kadowaki1986universal}, where $A$ is the coefficient of $T^2$ of low-temperature resistivity ($\rho=\rho_0+AT^2$) representing electron-electron scattering, and $\gamma_n$ is the linear coefficient in the heat capacity. Using $A=1.56(5)\times10^{-3}\ \mu\Omega$-cmK$^{-2}$ and $\gamma_n = 6.89(7)$ mJ-mol$^{-1}$K$^{-2}$, the value of $K_w$ is determined to be $32(1)\ \mu\Omega$-cmK$^{2}$J$^{-2}$mol$^{2}$ which exceeds $10\ \mu\Omega$-cmK$^{2}$J$^{-2}$mol$^{2}$ for heavy-fermion systems and indicates that it is a strongly correlated system \cite{miyake1989relation, takimoto1996relationship, kim2020strongly}.
Additionally, resistivity measurements at different applied magnetic fields were performed (shown in the top inset of \figref{Fig1}(c)) to ascertain the upper critical field, which will be addressed in the later paragraph. Furthermore, the carrier concentration was estimated by transverse measurement of resistivity $\rho_{xy}(H)$ at 20 K.  Fitting $\rho_{xy}(H)$ with a linear equation (shown in the bottom inset of \figref{Fig1}(c)) yields the Hall coefficient $R_\mathrm{H} = -2.9(1)\times10^{-10}\ \ohm$-mT$^{-1}$. The negative sign in $R_\mathrm{H}$ value indicates electrons as primary charge carriers. Using the equation $R_\mathrm{H} = -1/ne$, the carrier concentration $n$ is estimated to be $2.15(4)\times10^{28}\ \mathrm{m}^{-3}$ .

The temperature evolution of magnetization in zero-field-cooled-warming (ZFCW) and field-cooled-cooling (FCC) modes at 1.0 mT are shown in \figref{Fig2}(a) exhibiting a superconducting transition via a diamagnetic signal at an onset of $T_C=7.21(3)$ K. 
The separation between the ZFCW and FCC modes below $T_C$ is ascribed to significant flux-pinning effects. The magnetization loop at 1.8 K, illustrated in the inset of \figref{Fig2} (a), indicates type II superconductivity in Sc-Ti-V-Nb-Cu with de-pinning of vortices at $H_{irr}=6.2$ T. Furthermore, to determine the lower critical field $H_{C1}(0)$, isothermal magnetization in the low field range was measured from 1.8 to 7.0 K as shown in the inset of \figref{Fig2}(b). $H_{C1}$ for each curve was identified as the field value at which magnetization exhibits a deviation from the linear Meissner line. The temperature variation of the lower critical field, $H_{C1}(T)$, is accurately represented by the Ginzburg-Landau (GL) equation as;
\begin{equation}
    H_{C1}(T)=H_{C1}(0)\left[1-\left(\frac{T}{T_{C}}\right)^{2}\right].
    \label{eqn2:Hc1} 
\end{equation}
The best fit of $H_{C1}(T)$ using Eq. \ref{eqn2:Hc1} provides $H_{C1}(0) = 24.05(9)$ mT, as illustrated in \figref{Fig2}(b). In addition, the temperature evolution of magnetization (inset of \figref{Fig2}(c)) and resistivity (top inset of \figref{Fig1}(c)) were measured under different applied fields to estimate the upper critical field. The upper critical field, $H_{C2}(0)$, was estimated using the correlation of $T_C$ and the applied field, as $T_C$ shifts towards absolute zero when the applied field increases. The temperature variation of the upper critical field $H_{C2}(T)$ was fitted using the following GL equation, which is shown in \figref{Fig2}(c);
\begin{equation}
H_{C2}(T) = H_{C2}(0)\left[\frac{(1-t^{2})}{(1+t^2)}\right],
\label{eqn3:Hc2}
\end{equation}
where $t=\frac{T}{T_C}$ is the reduced temperature. The extrapolation of the fit yields $H_{C2}^{M}(0)=12.9(1)$ T from magnetization data and $H_{C2}^{\rho}(0)=15.9(2)$ T from resistivity data. The observed $H_{C2}(0)$ of both measurements is significantly high, indicating possible unconventionality.

Superconductivity in type-II materials can be suppressed by an external magnetic field exceeding the upper critical field through two main mechanisms: the orbital limiting effect and the Pauli paramagnetic effect. The orbital limiting field in the weak-coupling limit can be approximated by the WHH model as \cite{werthamer1966, helfand1966}:
\begin{equation}
    H_{C2}^{\mathrm{orb}}(0) = -\alpha T_C \left. \frac{dH_{C2}(T)}{dT}\right|_{T=T_{C}},
\label{eqn6:Hc2_orb}
\end{equation}
where $\alpha$ is 0.69 for dirty limit superconductors. The orbital limiting field $H_{C2}^{\mathrm{orb}}(0)$ is determined to be 6.7(5) T in the dirty limit using the slope value $\left|\frac{dH_{C2}(T)}{dT}\right|_{T=T_{C}}=1.35(1)$ T/K from the magnetization data. The Pauli paramagnetic field can be expressed as $H_{C2}^{\mathrm{P}}(0) = 1.84 T_C$ for BCS superconductors in the weak coupling limit \cite{chandrasekhar1962, clogston1962}. The $H_{C2}^{\mathrm{P}}(0)$ value is obtained as 13.2(6) T using $T_C=7.21(3)$ K. The Maki parameter, defined as $\alpha_\mathrm{M}=\sqrt{2} H_{c2}^{\mathrm{orb}}(0)/H_{c2}^{\mathrm{P}}(0)$, quantifies the relative strength of the orbital limiting effect compared to the Pauli paramagnetic effect \cite{maki1966}. The obtained value of $\alpha_\mathrm{M}$ is 0.75(8), indicating a small contribution of the Pauli paramagnetic field in pair-breaking.
\begin{figure}[ht] 
\includegraphics[width=0.9\columnwidth, origin=b]{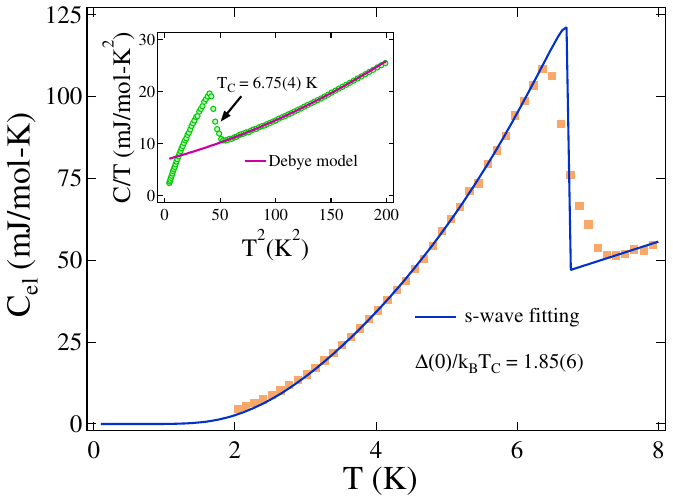}
\caption{\label{Fig3} (a) Electronic heat capacity with respect to temperature fitted with Eq. \ref{eqn11:s-wave}. The inset shows $C/T$ versus $T^2$ under zero magnetic field where the normal region is fitted to Eq. \ref{eqn7:Debye}.}
\end{figure} 

The Ginzburg-Landau coherence length, $\xi_{\mathrm{GL}}(0)$, is directly associated with the upper critical field $H_{C2}(0)$, through the equation $H_{C2}(0) = \frac{\phi_0}{2\pi \xi^2_{\mathrm{GL}}(0)}$, where $\phi_0 = 2.07\times10^{-15}$ T-m$^2$ denotes the magnetic flux quantum \cite{Tinkham}.
The value of $\xi_{\mathrm{GL}}(0)$ is determined to be 50.5(3) \AA, using $H_{C2}(0)=12.9(1)\ \mathrm{T}$. The penetration depth $\lambda_{\mathrm{GL}}(0)$ can be estimated using the following equation \cite{klimczuk2007physical}:
\begin{equation}
    H_{C1}(0) = \frac{\phi_0}{4\pi \lambda^2_{\mathrm{GL}}(0)}\left(\ln{\frac{\lambda_{\mathrm{GL}}(0)}{\xi_{\mathrm{GL}}(0)}+0.12}\right).
    \label{eqn5:lambda_GL}
\end{equation}
The value of $\lambda_{\mathrm{GL}}(0)$ is found to be 1559(15) \AA\  upon substituting $H_{C1}(0)$ and $\xi_{\mathrm{GL}}(0)$ into Eq. \ref{eqn5:lambda_GL}. The classification of superconductors into type I and type II is determined by the GL parameter ($\kappa_{\mathrm{GL}}$), defined as $\kappa_{\mathrm{GL}}=\frac{\lambda_{\mathrm{GL}}(0)}{\xi_{\mathrm{GL}}(0)}$. The value of $\kappa_{\mathrm{GL}}$ is calculated to be 30.8(5), which is significantly greater than $1/\sqrt{2}$, thus confirming type II superconductivity in Sc-Ti-V-Nb-Cu.

Heat capacity measurement at zero magnetic field was performed to examine the thermal properties of Sc-Ti-V-Nb-Cu, as shown in the inset of \figref{Fig3}. A significant jump in heat capacity at $T_C$ = 6.75(4) K confirms the existence of bulk superconductivity in this HEA. This $T_C$ value aligns with observations of magnetization and resistivity. The Debye-Sommerfeld model accurately describes $C(T)/T$ in the normal state:
\begin{equation}
    \frac{C(T)}{T} = \gamma_n +\beta_3T^2 +\beta_5T^4,
    \label{eqn7:Debye}
\end{equation}
where $\gamma_n$ represents the Sommerfeld coefficient, referring to the electronic contribution, while $\beta_3$ and $\beta_5$ refer to the phononic and anharmonic contribution, respectively. The fit to Eq. \ref{eqn7:Debye} provides the extrapolated values of $\gamma_n= 6.89(7)$ mJ-mol$^{-1}\mathrm{K}^{-2}$, $\beta_3 =0.057(1)$ mJ-mol$^{-1}\mathrm{K}^{-4}$, and $\beta_5=0.19(6)\ \mu$J-mol$^{-1}\mathrm{K}^{-6}$. The Debye temperature, $\theta_\mathrm{D}$ is related to $\beta_3$ via the following expression \cite{Kittel}:
\begin{equation}
    \theta_\mathrm{D} = \left(\frac{12\pi^4RN}{5\beta_3}\right)^{\frac{1}{3}},
    \label{eqn8:thetaD}
\end{equation}
where $R=8.31\ \mathrm{Jmol}^{-1}\mathrm{K}^{-1}$ is the universal gas constant, and $N$ represents the total number of atoms per formula unit, here $N=1$ for Sc-Ti-V-Nb-Cu. $\theta_\mathrm{D}$ is evaluated as 324(2) K using the values of $R$, $N$, and $\beta_3$ which lie in the elemental range. This value is larger than the one determined from resistivity measurements, which is also observed in \cite{landaeta2022conventional}. The density of states at the Fermi level, $D_C(E_\mathrm{F})$ for a non-interactive system is related to the Sommerfeld coefficient by the equation:
\begin{equation}
    \gamma_n = \left(\frac{\pi^2k^2_\mathrm{B}}{3}\right)D_C(E_\mathrm{F}),
    \label{eqn9:DOS}
\end{equation}
where $k_\mathrm{B}=1.38\times10^{-23}\ \mathrm{JK}^{-1}$ is the Boltzmann's constant. Using $\gamma_n= 6.89(7)$ mJ-mol$^{-1}\mathrm{K}^{-2}$, $D_C(E_\mathrm{F})$ is estimated to be 2.92(3) states/eV-f.u. To quantify the strength of the electron-phonon interaction, McMillan introduced the electron-phonon coupling constant $\lambda_{e-ph}$ given by the following equation in terms of $T_C$ and $\theta_\mathrm{D}$ \cite{mcmillan1968transition}:

\begin{equation}
\lambda_{e-ph} = \frac{1.04+\mu^{*}\ln{\left(\theta_\mathrm{D}/1.45T_{C}\right)}}{\left(1-0.62\mu^{*}\right)\ln{\left(\theta_\mathrm{D}/1.45T_{C}\right)}-1.04}.
\label{eqn10:e-ph}
\end{equation}
Here, $\mu^{*}$ refers to the screened Coulomb potential, which is taken as 0.13 for the inter-metallic compounds. $\lambda_{e-ph}$, is estimated to be 0.70(2), suggesting weak electron-phonon interaction in Sc-Ti-V-Nb-Cu.

The electronic heat capacity is calculated by subtracting the contribution of the lattice from the total heat capacity, expressed as $C_{el}(T)=C_{tot}(T)-\beta_3T^3-\beta_5T^5$. The normalized electronic heat capacity jump value is obtained as $\Delta C_{el}/\gamma_nT_c=1.59(2)$, which is slightly higher than the BCS value (1.43). The s-wave model of BCS superconductors effectively fits the low-temperature region of the electronic heat capacity for the normalized entropy S, as illustrated in \figref{Fig3}.
\begin{equation}
\frac{S}{\gamma_{n}T_{C}} = -\frac{6}{\pi^2}\left(\frac{\Delta(0)}{k_\mathrm{B}T_{C}}\right)\int_{0}^{\infty}\left\{f\ln{f}+(1-f)\ln{1-f}\right\}dy,
\label{eqn11:s-wave}
\end{equation}
where $f(\xi)=\left\{\exp\left(E(\xi)/k_\mathrm{B}T)\right)+1\right\}^{-1}$ represents the Fermi function, $E(\xi)=\sqrt{\xi^{2}+\Delta^{2}(t)}$ is the energy of normal electrons with integration variable $y=\xi/\Delta(0)$, and $\Delta(t)=\tanh{\left\{1.82(1.018((1/t)-1)^{0.51}\right\}}$ represents the BCS approximation of superconducting gap with reduced temperature $t=T/T_{c}$. The relationship between electronic heat capacity in the superconducting state and normalized entropy can be expressed as:
\begin{equation}
\frac{C_{el}}{\gamma_{n}T_{c}} = t\frac{d(S/\gamma_{n}T_{c})}{dt}.
\label{eqn12:gap}
\end{equation}
The superconducting gap value $2\Delta(0)/k_\mathrm{B}T_C=3.70(6)$ is obtained by fitting the electronic heat capacity in the superconducting region with Eq. \ref{eqn11:s-wave}. The value obtained is close to the BCS value (3.52) in the weak coupling limit, suggesting the conventional pairing of electrons \cite{padamsee1973quasiparticle}. 
\begin{figure}[ht]
\includegraphics[width=0.9\columnwidth, origin=b]{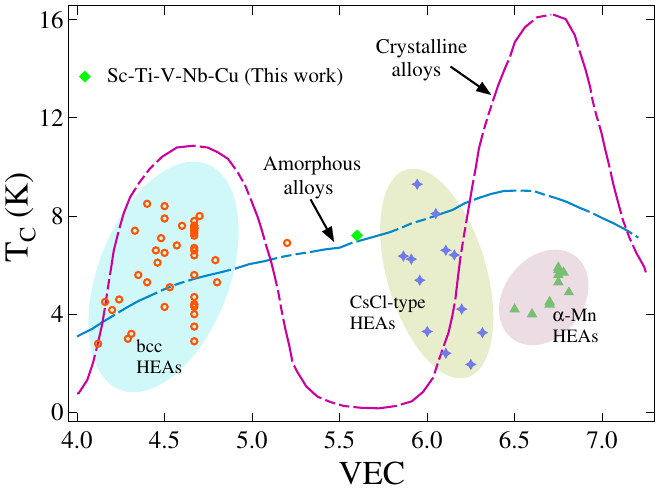}
\caption{\label{Fig4}$T_C$ vs. VEC plot for Sc-Ti-V-Nb-Cu where dotted pink and blue lines are the trends for crystalline and amorphous alloy, respectively. For reference, $T_C$ and VEC of other HEAs with different crystal structures are also shown \cite{matthias1955empirical, collver1973superconductivity, kitagawa2020cutting, jangid2024superconductivity}.}
\end{figure}

\label{subsec4}
\figref{Fig4} illustrates the correlation between $T_C$ and the valence electron concentration. This analysis includes comparative data from various sources on crystalline metals, amorphous metals, body-centered cubic (bcc), CsCl-type, and a-Mn high-entropy alloys \cite{matthias1955empirical, collver1973superconductivity, kitagawa2020cutting, jangid2024superconductivity}. In particular, the VEC dependence of $T_C$ for Sc-Ti-V-Nb-Cu is equivalent to amorphous alloys, making this HEA an excellent system for studying superconductivity at the edge of crystalline and amorphous alloys.

The experimentally obtained parameters, residual resistivity ($\rho_0$), carrier concentration ($n$), and Sommerfeld coefficient ($\gamma_n$), are used to assess the electronic parameters using a series of equations. The effective mass ($m^*$) is related to $\gamma_{n}$ and $n$ by the equation $\gamma_{n} = \left(\frac{\pi}{3}\right)^{2/3}\frac{k_\mathrm{B}^{2}m^{*}n^{1/3}}{\hbar^{2}}$, where $\hbar=1.05\times10^{-34}$ Js is the reduced Planck's constant. The effective mass $m^*$ is evaluated as $15.45(5)\ m_e$ after substituting the values of $\gamma_n$ and $n$ (from Hall measurement). The Fermi velocity $v_\mathrm{F}$, is calculated to be $v_\mathrm{F}=0.64(4)\times10^5 \ \mathrm{ms}^{-1}$ by the expression $n=\frac{1}{3\pi^2}\left(\frac{m^*v_\mathrm{F}}{\hbar}\right)^3$. 
The mean free path ($l$) is determined to be $74.6(4)$ \AA\  using the equation $l = \frac{3\pi^2\hbar^3}{e^2\rho_0m^{*2}v^2_{\mathrm{F}}}$ and $\rho_0 = 22.02(1)\ \mu\Omega$-cm. The BCS coherence length ($\xi_0$) can be determined by the expression $\xi_0=\frac{0.18\hbar v_{\mathrm{F}}}{k_\mathrm{B}T_C}$. Substituting the values of $v_{\mathrm{F}}$ and $T_C$ yields $\xi_0=122(8)$ \AA. Based on the ratio of the BCS coherence length to the mean free path ($\xi_0/l$), superconductors can be categorized into clean and dirty limit superconductors. In the case of Sc-Ti-V-Nb-Cu, the ratio $\xi_0/l=1.64(6)$ is greater than 1, suggesting that it is a dirty limit superconductor.

Uemura \textit{et al.} developed a scheme for classifying conventional and unconventional superconductors employing the ratio of the superconducting transition temperature ($T_C$) and the Fermi temperature ($T_\mathrm{F}$) \cite{uemura1988systematic, uemura1991basic}. Unconventional superconductors have $0.1 \geq T_C/T_\mathrm{F} \geq 0.01$, whereas conventional superconductors have $T_c/T_\mathrm{F}\leq 0.01$. For a 3D non-interacting system, the Fermi temperature is expressed as \cite{hillier1997classification}:
\begin{figure}[ht]
\includegraphics[width=0.9\columnwidth, origin=b]{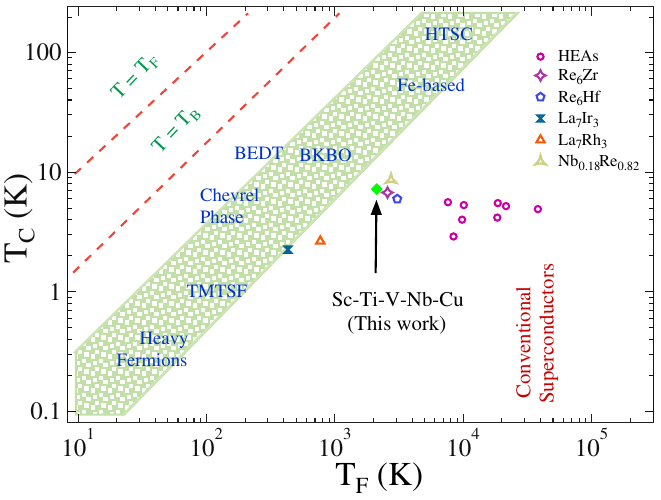}
\caption{\label{Fig5}Uemura classification of superconductors by $T_C$ and Fermi temperature $T_{\mathrm{F}}$. The cream-colored band represents unconventional superconductors, and Sc-Ti-V-Nb-Cu lies close to this band.}
\end{figure}

\begin{table}[ht]
\caption{Superconducting and normal state parameters of Sc-Ti-V-Nb-Cu.}
\label{Tab2}
\begin{center}
\begingroup
\setlength{\tabcolsep}{8
pt}
\begin{tabular}{c c c} 
\hline\hline
Parameter & Unit & Value  \\ [1ex]
\hline
VEC& & 5.6 \\  
RRR&&2.44\\
$T_{C}$& K& 7.21(3)\\            
$H_{C1}(0)$& mT& 24.05(9)\\                       
$H_{C2}(0)$& T& 12.9(1)\\
$H_{C2}^\mathrm{P}(0)$& T&13.2(6)\\
$H_{C2}^{Orb}(0)$& T& 6.7(5)\\
$\xi_\mathrm{GL}(0)$& \AA & 50.5(3)\\
$\lambda_\mathrm{GL}(0)$& \AA & 1559(15)\\
$k_\mathrm{GL}$& &30.8(5)\\
$\gamma_{n}$&  mJmol$^{1}$-K$^{2}$& 6.89(7)\\
$\theta_\mathrm{D}$& K& 324(2)\\
$\lambda_{e-ph}$& &0.70(2)\\
$\Delta C_{el}/\gamma_nT_C$ & &1.59(2)\\
$2\Delta(0)/k_\mathrm{B}T_C$& &3.70(6)\\
$\xi_{0}/l_{e}$& &  1.64(6)\\
$v_{\mathrm{F}}$& 10$^{5}$ ms$^{-1}$& 0.64(4)\\
$n$& 10$^{28}$m$^{-3}$& 2.15(4)\\
$T_{\mathrm{F}}$& K&2117(54)\\
$m^{*}$/$m_{e}$&  & 15.45(5)\\
[1ex]
\hline\hline
\end{tabular}
\endgroup
\end{center}
\end{table}
\begin{equation}
    k_\mathrm{B}T_\mathrm{F} = \frac{\hbar^2}{2m^*}\left(3\pi^2n\right)^{\frac{2}{3}}.
    \label{eqn18:Fermitemp}
\end{equation}
Upon substituting the values of $m^*$ and $n$, Eq. \ref{eqn18:Fermitemp} yields $T_\mathrm{F}=2117(54)$ K and $T_C/T_\mathrm{F}=0.0034(5)$ positioning it close to the unconventional band as illustrated in \figref{Fig5}. Table \ref{Tab2} summarizes all normal and superconducting parameters for Sc-Ti-V-Nb-Cu.

In conclusion, we have successfully synthesized a 3d-dominated, lightweight equiatomic high-entropy alloy, Sc-Ti-V-Nb-Cu, which adopts a body-centered-cubic (bcc) structure. Our detailed investigation, including magnetization, electrical resistivity, and specific heat measurements, explores its normal and superconducting properties. This HEA is a type-II bulk superconductor with a transition temperature of 7.21(3) K. The upper critical field, derived from both magnetization and resistivity data, approaches the Pauli paramagnetic limit, suggesting possible unconventional behavior.
Heat capacity measurements indicate an isotropic superconducting gap with s-wave pairing symmetry. The Kadowki-Woods ratio classifies this HEA as a strongly correlated system. These combined properties make this lightweight equiatomic HEA an ideal platform for investigating unconventional superconductivity. Its low mass density, moderate transition temperature, high critical field, and potential for thin-film fabrication render it promising for various device applications. To gain deeper insights into the superconducting pairing mechanism and the impact of disorder, further microscopic studies and theoretical electronic structure analyses are imperative.

\section*{ACKNOWLEDGEMENT} 
R.P.S. acknowledges the SERB Government of India for the Core Research Grant No. CRG/2023/000817.
\pagestyle{plain}
\addcontentsline{toc}{chapter}{Bibliography}
\bibliographystyle{apsrev}
\bibliography{reference}

\end{document}